\documentclass{amsart}

\usepackage[english]{babel}
\usepackage[a4paper, total={7in, 10in}]{geometry}

\usepackage{amsmath}
\usepackage{chemformula}
\usepackage{graphicx}
\usepackage{xcolor}
\usepackage{amsaddr}
\usepackage{amsmath,amsfonts,amssymb}
\usepackage{algorithm}
\usepackage{algpseudocode}
\usepackage{float}
\usepackage{upgreek}
\usepackage{cite}

\newcommand{\R}{\mathbb{R}}
\newcommand{\N}{\mathbb{N}}

\newcommand{\Z}{\mathbb{Z}}

\newcommand{\compactness}{\gamma}
\newcommand{\diameter}{d_\mathrm{max}}
\newcommand{\minDiameter}{d_{\mathrm{min}}}
\newcommand{\aspectRatio}{A_{\mathrm{R}}}
\newcommand{\sphericity}{\Psi}
\newcommand{\E}{\mathbb{E}}

\newcommand{\grainSize}{A_\mathrm{cross}}
\newcommand{\chordLength}{c}
\newcommand{\elongation}{\eta}
\newcommand{\orientation}{\alpha}

\newcommand{\numSeeds}{\nu}

\newcommand{\TGen}{T_\mathrm{3D}}
\newcommand{\TReal}{T_\mathrm{2D}}

\newcommand{\GNet}{G_\mathrm{net}}

\newcommand{\DifTesRep}{\mathcal{T}}

\newcommand{\numberGrainsGT}{NG^\mathrm{gt}}
\newcommand{\numberGrainsGenerated}{NG^\mathrm{gen}_e}

\newcommand{\round}{\xi}
\newcommand{\argmin}{\mathop{\mathrm{arg\,min}}}

\newcommand{\consideredNeighbours }{\kappa}

\definecolor{ulmgruen}{rgb}{0.3372,0.6667,0.1098}

\author{Lukas Fuchs$^1$, Orkun Furat$^1$, Donal P. Finegan$^2$, Jeffery Allen$^2$, \\Francois L.E. Usseglio-Viretta$^2$, Bertan Ozdogru$^2$, \\Peter J. Weddle$^2$, Kandler~Smith$^2$, Volker Schmidt$^1$}
\address{$^1$ Ulm University, Institute of Stochastics, Helmholtzstraße 18, 89069 Ulm, Germany}
\address{$^2$ National Renewable Energy Laboratory, 15013 Denver W Parkway, Golden, CO 80401, USA}

\title[Generating multi-scale NMC particles]{Generating multi-scale NMC particles with radial grain architectures using spatial stochastics and GANs}

\begin{document}

\begin{abstract}

Understanding structure-property relationships of Li-ion battery cathodes is crucial for optimizing rate-performance and cycle-life resilience. However, correlating the morphology of cathode particles, such as in NMC811, and their inner grain architecture with electrode performance is challenging, particularly, due to the significant length-scale difference between grain and particle sizes. Experimentally, it is currently not feasible to image such a high number of particles with full granular detail to achieve representivity. A second challenge is that sufficiently high-resolution 3D imaging techniques remain expensive and are sparsely available at research institutions. To address these challenges, a stereological generative adversarial network (GAN)-based model fitting approach is presented that can generate representative 3D information from 2D data, enabling characterization of materials in 3D using cost-effective 2D data. Once calibrated, this multi-scale model is able to rapidly generate virtual cathode particles that are statistically similar to experimental data, and thus is suitable for virtual characterization and materials testing through numerical simulations. A large dataset of simulated particles with inner grain architecture has been made publicly available.

\smallskip

\noindent
\textbf{Keywords: NMC811,  multi-scale model, stereology, tessellation, GAN, virtual cathode particle, grain architecture} 

\end{abstract}
\maketitle
\newpage

\section{Introduction}

Improving Li-ion battery energy-density, power-density, and cycle-life is intricately linked to addressing the climate crisis by enabling transportation electrification and increasing storage capabilities for renewable energy sources. The tremendous progress in Li-ion battery performance, achieved over the past 30 years, has largely stemmed from fine tuning electrode microstructures and electrolyte compositions, without much change in the active material elemental composition~\cite{D15}. Yet, opportunities still exist for controlling cathode sub-particle microstructure to enhance performance~\cite{TWYCCFLPKAUPBDSCWWR22,ALLEN2021230415}. For example, frabricating radially oriented and elongated grains within  LiNi$_x$Mn$_y$Co$_z$O$_2$ (NMC$xyz$) cathode particles has been shown to significantly increase rate capability as compared to cathode particles that have no preferential grain orientation or internal architecture~\cite{TWYCCFLPKAUPBDSCWWR22,RPYSSLYDMA19,KPSNLKKYS20,XHJWZXHYDS19}.  It is also expected that by designing cathode grain microstructures, chemo-mechanically induced fracture, which perpetuates loss-of-active-material and eventually leads to battery capacity fade, can be reduced~\cite{ALLEN2021230415}.  The present paper seeks to improve the understanding of structure-property relationships between particle-level and grain-level microstructures and observed battery performance.  This is achieved by developing a tool that can generate representative microstructures from experimental data.  These realistic microstructures are essential in experiment data interpretation and are required for physics-based models that seek to explore particle- and grain-level physics that ultimately determine battery performance and longevity.

There is a need in the community for a tool that can generate representative microstructures for mesoscale battery models to assist in mapping realistic microstructural information to cell-level performance.  Physics-based models have been used to help accelerate Li-ion battery electrode development. For example, physics-based models can be used to optimize designs for fast-charge performance~\cite{Colclasure_2019,KWGMC23,RSCSTBA21,MCS20}.  Effective optimization using battery models requires relatively fast computation times.  To achieve fast computation times, physics-based models typically abstract detailed electrode-level, particle-level, and grain microstructures to ``effective'' parameters that approximate these intricacies~\cite{Doyle_1993}.  Examples of these effective parameters include: secondary-particle diameter, specific surface area, solid-phase diffusion coefficients, and electrolyte tortuosity.  While these fast electrochemical models have shown great promise, there is a disconnect between the optimized effective parameters and the underlying complex microstructures that can achieve these performance metrics.  

Moreover, geometric complexities not typically included in physics-based models, such as grain orientation, grain size, and secondar particle surface area, can strongly influence cell aging~\cite{TYFCLWBCDWTESADQDTJ22,TWYCCFLPKAUPBDSCWWR22}.  For example, secondary particle cracking has a complex relationship with capacity fade--cracks reduce solid-phase diffusion lengths and increase specific surface area~\cite{PCMS23}, but can isolate active material and introduce unprotected surfaces for side reactions~\cite{MLXHXSSWNDLZ18,MWXZWBSLYXLYHL19,KCKHLL21}).  Additionally, grain orientation, size, and shape greatly influence the apparent solid-phase diffusion coefficient~\cite{USSEGLIO_2019}. 

To bridge the gap between optimized effective parameters, identified in physics-based models and real microstructures, researchers have developed ``mesoscale'' models that focus on particle- and grain-level microstructures and associated electro-chemo-mechanic physics~\cite{CARFSSWSX24,TWBK21,XYYLCLLZ19,XVSLZ17,MWXZWBSLYXLYHL19,BZLSX20}.  However, generating statistically representative, realistic particle and grain microstructures as inputs for these mesoscale models is challenging~\cite{RMBTNG16,MBZK14}.  Consequently researchers often approximate geometries with idealized shape, size, and orientations~\cite{SP22,GCWK12,TWBK21,TWHBK23,XYYLCLLZ19,ZCCL20,XVSLZ17,MWXZWBSLYXLYHL19,BZLSX20,SSSZLK16}.  

In the present paper, a ``multi-scale'' particle model is developed to generate representative secondary NMC811 particles comprised of representative grains (primary-particles).  The multi-scale approach can be subdivided into two distinct models that can generate virtual morphologies that statistically resemble secondary-particles and grain architectures.  A similar two-part, multi-length-scale approach was used previously~\cite{FuratMultiScale}.  As in this study, the outer secondary-particle shell is generated using a random field on the unit sphere~\cite{SHfeinauer2015structural}, and the inner grain architecture is generated using a random tessellation. Roughly speaking, common (distance-based) tessellations can be considered as parametric partitions of Euclidean space, such as the Voronoi or Laguerre diagram~\cite{doi:10.1137/Voronoi}.
Unlike the previous study~\cite{FuratMultiScale}, the proposed grain architecture model is not based on Laguerre tessellations~\cite{Lautensack_Zuyev_2008}, but on a more general class. Furthermore, in contrast to previous approaches, a neural network is implemented to generate the parameters of the tessellations. Importantly, this allows for a novel (gradient descent-based) fitting procedure of the tessellation model, enabling the generation of 3D grain architectures solely by using 2D planar section information.

Several neural-network-based approaches for 2D-3D (re-)construction approaches have been explored previously~\cite{SliceGAN,pix2vox,fuchs2024using}. In contrast to the present model, most of these black-box methods are based on upsampling~\cite{Upsampling} or transposed convolution~\cite{DeConv} techniques to generate  discrete 3D voxel representations from 2D pixel images (i.e., slices or projections). These generated 3D voxel representations do not have any underlying constraints and thus are able to generate any morphology at the expense of interpretability and computational efficiency. 

In contrast to other 2D-to-3D generation methods, the present method does not solely rely on neural networks and the generation of a 3D voxel-based representation. Instead, a neural network and  a Poisson point process~\cite{chiu2013stochastic,PoissonLast} is used to generate (continuous) tessellation parameters~\cite{petrich2021efficient,buze2024anisotropic}. More precisely, the parameters of the random tessellation model, i.e., the parameters of the neural network and the point process, are fit by a generative adversarial network (GAN)-based~\cite{GANOverview} approach, allowing for the simulation of realistic 3D grain architectures.  
These virtual grain architectures are then sliced into several 2D images and passed to the so-called discriminator to be compared with measured grain architectures. The output of the discriminator is then used to update the parameters of the grain architecture model to generate more realistic 3D grain architectures.  After training, the generated 3D structures are statistically similar to ones measured experimentally. This enables characterization of 3D features like the distribution of morphological features of grains, by using only 2D image data, bridging a major systematic barrier in characterizing particle architectures in 3D. These simulated 3D grain architectures, i.e., continuous tessellations, are especially suitable for simulation applications~\cite{ALLEN2021230415,CARFSSWSX24,XYYLCLLZ19,BZLSX20,BMWZBHBS22, FZC21}. Furthermore, the fitted parameters of the random tessellation model are interpretable, and by modifying these parameters, similar structures with novel hypothetical structural characteristics can be generated and virtually tested using numerical simulations of electrochemical performance~\cite{neumann2020VirtualTesting, westhoff2018VirtualTesting,CARFSSWSX24} or cracking behavior~\cite{ALLEN2021230415}.
Alongside this manuscript, a data set of simulated NMC811 particles with inner grain architecture is made publicly available. 

\section{Materials;  acquisition,  preprocessing and analysis of image data}

\subsection{Overview of experimental approach} 

To supply sufficient high-quality image data on particle and grain architectures, a multi-modal approach was deployed. Electron backscatter diffraction (EBSD) was used, as previously described 
in~\cite{QuinnEBSD}, to facilitate grain and grain boundary segmentation, as well as populate the segmented grains with local crystallographic orientation information. Note, there is a distinction between crystallographic and morphological grain orientation; crystallographic orientation refers to the orientation of the $c$-axis of the NMC811 crystal, whereas morphological grain orientation refers to the direction of the major axes of the best fitting elongated spheroid shapes of the grains with respect to the particle center. Conducting EBSD on a small number of particles provides a view of hundreds of grain planar sections. However, extending EBSD to 3D via focused ion beam (FIB)-EBSD and acquiring morphological data on hundreds of full particles is not yet practical experimentally. Instead, the (secondary-) particle morphological information, or outer shell, was captured in detail for many particles by X-ray nano-computed tomography (nano-CT) at the cost of a lack of inner grain architecture information. The outer shell and sub-particle grain architectures are independent and therefore must each be quantified with a sufficient number of samples to be representative. The combination of 2D EBSD and 3D nano-CT facilitated detailed information on the sub-particle grain features and full particle outer shell morphological detail, respectively, with sufficient volume to achieve representivity in both.

\subsubsection{Materials} 

A pristine sample of NMC811 electrode was used for this work. The electrode consisted of 96~wt\% Targray NMC811, 2~wt\% Timcal C-45, and 2~wt\% Solvay 5130 PVDF binder. The Al foil was 20~$\upmu$m thick, and the coating thickness was 58~$\upmu$m with a porosity of 33\% and areal capacity of 3.07~mAh~cm$^{-2}$.  These cathodes have been extensively studied in fast-charge applications for electric vehicles~\cite{TWYCCFLPKAUPBDSCWWR22,TYCCGLYWWDBSDETDTPJ21,CCCTDSEDTPJ21,TYFCLWBCDWTESADQDTJ22}. The preferential radial orientation of these particles is argued to significantly increase these cathodes rate performance and cycle-life resiliency~\cite{TYFCLWBCDWTESADQDTJ22}.

\subsubsection{X-ray nano-computed tomography (nano-CT)} 

For X-ray nano-CT imaging, cylindrical pillars of ca. 90~$\upmu$m diameter were prepared using a micro-machining laser ablation approach described previously in~\cite{BaileyLaser}. The pillars were then imaged in a Zeiss Xradia Ultra 810 X-ray nano-CT system in Large Field of View (LFOV) absorption mode with binning 2, giving a voxel size of 128~nm. The field of view was 64~$\upmu$m~$\times$~64~$\upmu$m.  1601 images were acquired for the reconstruction of each tomogram.

\subsubsection{SEM and EBSD imaging} \label{sec:EBSD_imaging}

The NMC811 cathode sample was argon-milled with a JEOL CP ion beam planar section polisher (JEOL, USA). This provided a wide (ca.~2~mm) smooth planar section of the NMC811 electrode. Scanning electron microscopy (SEM) and EBSD images were taken on several planar sectioned particles using an FEI Nova NanoSEM 630 equipped with an EBSD detector (EDAX, USA). EBSD data was collected using step sizes of 50~nm rastered across the surface of particle planar sections. EBSD data was processed with OIM Analysis v.8 (EDAX, USA). Diffraction patterns were fit to a trigonal crystal system (space group R-3~m) with $a$ = $b$ = 2.875~\AA\ and $c$ = 14.248~\AA\ to obtain the orientation of the crystal at each, where $a,b,c$ denote the side lengths of the rhombohedral unit cell associated with the crystal's atomic lattice. The software produced text files containing a spatially resolved confidence index, image quality (IQ), and Bunge-Euler angle data. These data were then converted into images, where individual grains are labeled separately.

\subsection{Preprocessing of nano-CT  data}\label{sec.pre.pro}

The NMC811 particles' outer shells are modeled by a random field on the sphere. Fitting the random field model requires preprocessing the nano-CT data.

\subsubsection{Segmentation and labeling} \label{sec:segmentation_and_labeling}

\begin{figure}[H]
        \centering
        \includegraphics[width=0.9\textwidth]{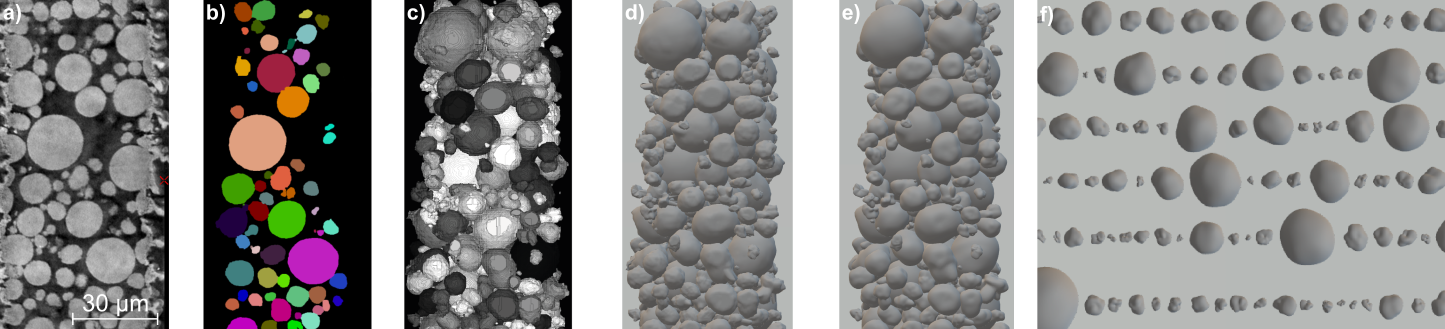}
        \caption{Subsequent data-preprocessing steps for  modeling the outer shell of NMC particles: a) Planar section of 3D nano-CT data. b) Segmentation of the planar section shown in a). c) 3D rendering of voxel-based particle representation. d) Star-shaped representation of particles displayed in c). e) Spherical harmonics-based representation of particles shown in d). f) simulated particles generated by the stochastic  outer shell model (explained in Section~\ref{sec:outer_shell} below). 
        }
        \label{fig:outer_shell_process}
    \end{figure}

Initially, the 3D grayscale CT image, see Figure~\ref{fig:outer_shell_process}a, undergoes segmentation and labeling. This process includes convolving the grayscale image with a discrete 3D Gaussian kernel with standard deviation $\sigma=1.8$, followed by Otsu thresholding~\cite{Otsu} and labeling of the resulting binary 3D image using the watershed algorithm~\cite{watershed}. Subsequently, disconnected components of the watershed-labeled image are separated, and small (volume $V<10$ voxels) or non-spherical (sphericity $\Psi<0.5$, see Eq.~(\ref{eq::sphericity}) for a formal definition) components are merged with their corresponding neighboring components if they exist; otherwise, they are neglected to remove artifacts. Finally, components intersecting the boundary of the sampling window are removed to minimize edge effects. This procedure yields $N=1590$ labeled particles in voxel representation, as shown in Figures~\ref{fig:outer_shell_process}b and~\ref{fig:outer_shell_process}c.

\subsubsection{Spherical harmonics representation}

These voxel-based particles are assumed to be star-shaped such that their so-called star point coincides with their  centers of masses. Thereby, a star-shaped domain is one where every line segment from the star point to a boundary point lies entirely inside the domain. Therefore, a particle with center of mass $c\in \R^3$ can be represented by a radius function $P:S^2\to\R_+=[0,\infty)$ on the unit sphere $S^2=\{x \in \R^3: |x|=1\}$ given by
\begin{align}
    P(u)=\max\{r \in \R_+: \text{voxel }\round (ru+c) \text{ belongs to the particle}\}, ~~~\qquad\text{for each } u\in S^2,\label{eq::particle_representation_radius_function}
\end{align}
where $\round(x)\in \Z^3$ denotes the grid point that is closest to some point $x\in \R^3$. For a visual validation of the star-shape assumption, see Figure~\ref{fig:outer_shell_process}d.

To stochastically model the star-shaped particles, the particles radius is modeled as a function by means of random fields on $S^2$. Realizations of these models can be considered to be radius functions that describe the outer shell of particles. Therefore, we use so-called (real-valued) spherical harmonic functions $Y_{\ell m}\colon S^2 \to \R$ for $\ell \in \N_0=\{0,1,\ldots\}$ and $m\in \Z=\{\ldots,-1,0,1,\ldots\}$ with  $|m|\leq \ell$, which are given by
\begin{align}
    Y_{\ell m}(u)=\begin{cases}
 \displaystyle \sqrt{2}\sqrt{\frac{2\ell +1}{4 \pi}\frac{(\ell -|m|)!}{(\ell+|m|)!}} P^{|m|}_\ell(\cos\theta(u))\sin(|m|\phi(u)), & \text{if } m <  0, \\
 \displaystyle
  \sqrt{\frac{2\ell +1}{4 \pi}} P^{m}_\ell (\cos\theta(u)),  & \text{if } m = 0,\\
  \displaystyle\sqrt{2}\sqrt{\frac{2\ell +1}{4 \pi}\frac{(\ell-m)!}{(\ell+m)!}} P^{m}_\ell (\cos\theta(u))\cos(m\phi(u)), & \text{if } m > 0,
\end{cases}\qquad \text{for each }u \in S^2,
\label{def.sph.har}
\end{align}
where $P^{m}_\ell \colon [0,1] \to [0,1]$ are the Legendre polynomials and $\theta(u)\in[0,\pi)$, $\phi(u)\in[0,2\pi)$ are the polar and the azimuthal angles of $u\in S^2$, respectively. These functions form an orthonormal basis of the Hilbert space $L_2(S^2)$ of square-integrable functions\footnote{A function $P\colon S^2 \to \R$ is called square-integrable if $ \int_{S^2} P^2(u) \,\mathrm{d}u < \infty$.} on the unit sphere $S^2$~\cite{atkinson2012spherical}. Thus, loosely speaking, the surface of each star-shaped particle can be expressed as a linear combination of spherical harmonic functions. As the order of these spherical harmonic functions increases (i.e., the value of $\ell$), they contribute to capturing more detailed and rougher surface features. Since the considered CT data has a finite resolution, it is not meaningful to consider significantly high orders.
Therefore, in the following only spherical harmonic basis functions with $\ell \leq L = 9$ are considered.
This results in a spherical-harmonics representation $ \widetilde{P}:S^2\to\R_+$ of the radius function $P$, given by
\begin{align}\label{lab.peh.uhh}
    \widetilde{P}(u)=\sum_{\ell=0}^L\sum_{m=-\ell}^l a_{\ell m}Y_{\ell m}(u), \qquad \text{for each }u\in S^2,
\end{align}
where $a_{\ell m}$ are the $(L+1)^2=100$ coefficients of this basis representation. The computation of these coefficients for a given star-shaped particle's radius function $P$, and therefore its approximation in terms of the spherical harmonics basis up to order $L$, is done by least squares regression.
More precisely, the least squares fit between $\widetilde{P}(Ru)$ and $P(u)$ is computed regarding 642 equidistant evaluation points $u\in S^2$ on the unit sphere, resulting in 642 equations with the 100 variables $a_{\ell m}$. Here, $R\in \R^{3\times 3}$ denotes a rotation matrix, uniformly chosen at random, which helps to mitigate any anisotropy in the data arising from the voxel-based representation of the CT data or the segmentation procedure. For a visual inspection of the  quality of this fit, see Figure~\ref{fig:outer_shell_process}e.

\subsection{Morphology of 2D grain architectures}
\label{sec:descriprion_of_orientation}
\begin{figure}[H]
        \centering
        \includegraphics[width=.8\textwidth]{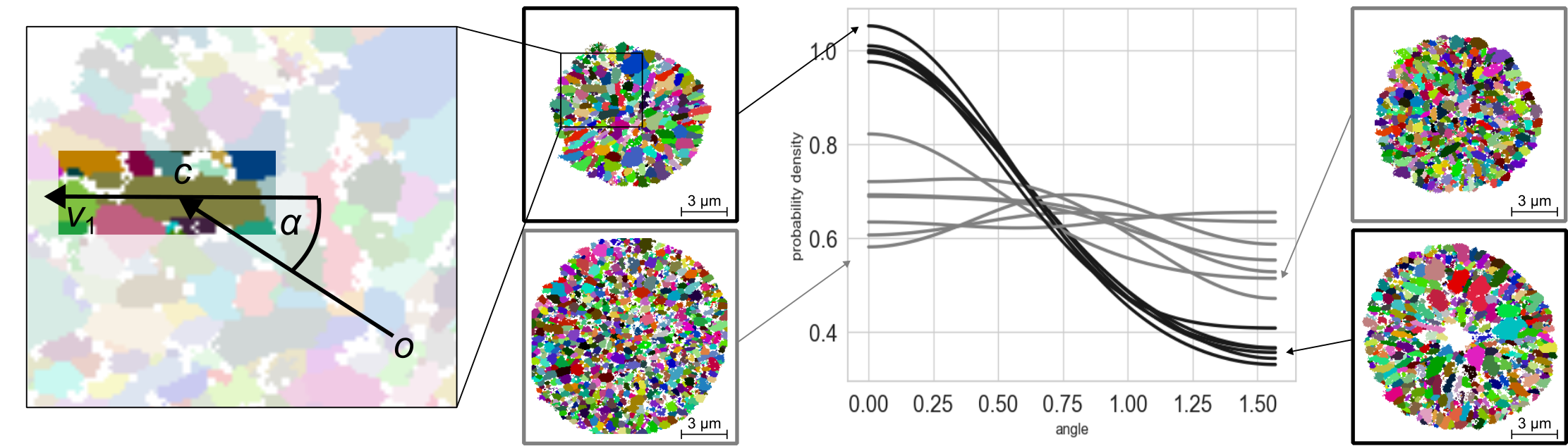}
        \caption{Orientation of grain planar sections: The orientation angle $\orientation$ of a grain is computed via its first principal component $v_1$ and its center $c$ (first column from the left). The angle distribution of each EBSD planar section is determined by  kernel density estimation, using symmetric boundary conditions. The curves in black correspond to retained data, whereas the gray curves correspond to neglected data  (third column). Two exemplary grain architectures of retained and neglected planar section data are shown (second and fourth columns).}
        \label{fig:grain_orientation}
    \end{figure}

As mentioned in Section~\ref{sec:EBSD_imaging}, the text data produced by the EBSD measurement was first converted into image data of labeled grains, as shown in Figure~\ref{fig:grain_orientation}. These 2D image data resulting from EBSD measurement are from now on referred to as 2D EBSD (planar section) data. The grain architectures observable in these data show a non-uniform morphological orientation, i.e., the grains morphologies exhibit a preferred orientation towards the center of the particle planar section, which is identified as the origin $o=(0,0)\in \R^2$ of the coordinate system. Note that morphological orientation described here is distinct from crystallographic orientation. The proposed model should be able to reflect this behavior, thus, the orientation of individual 2D grains has to be captured quantitatively. For this purpose, the principal components $v_1,v_2 \in \R^2$ of the pixel positions associated with a grain, along with their corresponding eigenvalues $e_1 \geq e_2 > 0$ are utilized. 
The first principal component $v_1$ is a direction that maximizes the (empirical) variance of the pixel positions~\cite{jolliffe2016principal}. In other words, it describes the primary orientation of the grain. The orientation $\orientation\in [0,\frac{\pi}{2}]$ of a grain is thus given by the smallest angle between the grain's center $c$ (in the planar section) and $v_1$, i.e.,
\begin{align}
\orientation &= \arccos\left( \frac{|v_1c^\top|}{|v_1||c|}\right),\label{eq:orientation}
\end{align}
see Figure~\ref{fig:grain_orientation}.
For grains with $\orientation \approx 0$ the principal component $v_1$  points towards the origin, whereas an orientation $\alpha = \pi/2$ corresponds to principal components that are orthogonal to the direction pointing to the origin. Figure~\ref{fig:grain_orientation} shows the grain orientation distribution per 2D EBSD image. Furthermore, exemplary planar sections are presented in Figure~\ref{fig:grain_orientation}. Notably, for some planar sections, the corresponding EBSD data (outlined with gray boxes in Figure~\ref{fig:grain_orientation}) exhibits no preferred grain orientation. It is conjectured that these images correspond to planar sections located farther away from the particle center as compared to those highlighted with a black box, i.e., the assumption that the particle center is located at the origin of the planar section observed in the EBSD image is not valid. However, the stereological model fitting approach for the inner grain architecture model, introduced in Section~\ref{sec:tess_fit}, relies on image data of planar sections passing through the particle center, since in such images, grains in each radial distance from the particle center can be observed, a property not present in planar sections at other locations. Consequently, EBSD planar sections represented by gray curves in Figure~\ref{fig:grain_orientation} are excluded from the subsequent model fitting procedure.

\section{Stochastic multi-scale model}

In this section, the stochastic  models for both the outer shell and the inner grain architecture, along with their corresponding fitting procedures, are presented. The choice of a two-scale modeling approach is driven by the presumed minimal correlation between the outer shell of NMC811 particles and their inner grain architecture. Initially, the outer shell structure is modeled on a coarser length scale (by means of nano-CT data), before proceeding to the modeling of the inner 3D grain architecture, which is observable on a finer length scale (by means of 2D EBSD data). 

\subsection{Stochastic outer shell model}
\label{sec:outer_shell}

\begin{figure}[H]
\begin{minipage}{.38\textwidth}
    \centering\vspace{-0.3cm}
    \includegraphics[width=\textwidth]{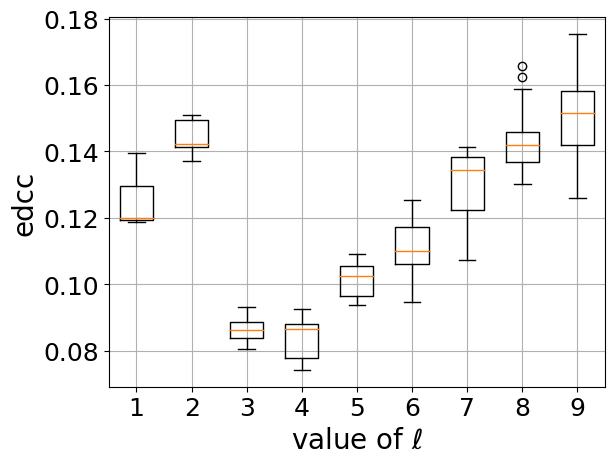}
\end{minipage}
\hspace{10pt}
\begin{minipage}{.58\textwidth}
    \centering
    \vspace{-15pt}
    \includegraphics[width=\textwidth]{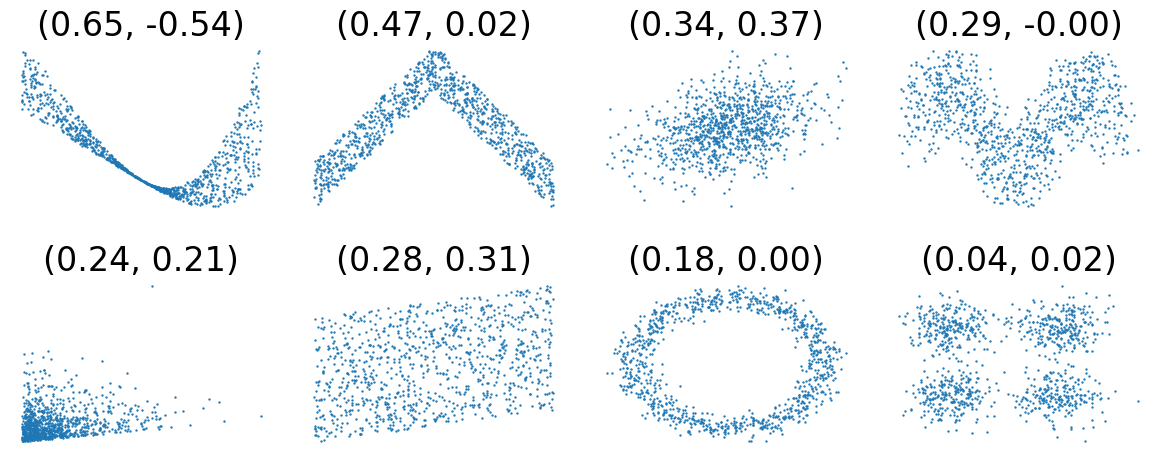}
\end{minipage}
    \caption{Empirical distance correlation coefficient (edcc) between  $a_{00}$ and $a_{\ell m}$ of nano-CT data (left), and plots of further exemplary data with corresponding empirical correlation coefficients (right), where the values above the plots are the edcc and the Pearson correlation coefficient, respectively.  Values of  edcc that are close to 0  and 1 indicate weak and strong dependence, respectively.}
    \label{fig:corr_coef}
\end{figure}

After the preprocessing of nano-CT data explained in Section~\ref{sec.pre.pro}, 
the next step towards stochastic outer shell modeling is to describe the distribution of the coefficients $a_{\ell m}$ appearing in Eq.~\eqref{lab.peh.uhh} by appropriately chosen random variables $Z_{\ell m}$. Due to the assumed isotropy of the underlying image data, the random variables $Z_{\ell m}$ and $Z_{\ell m'}$ are supposed to be identically distributed for any $m,m'$. For $(\ell,m) \neq (\ell',m') $ with $\ell,\ell' \neq 0$,
the random variables $Z_{\ell m}$ and $Z_{\ell' m'}$ can be assumed to be independent, see \cite{Lang_2015}. Furthermore, to justify the assumption of independence of  $Z_{0 0}$ and $Z_{\ell m}$ for $\ell,m\neq 0$, the empirical distance correlation coefficient (edcc)~\cite{distanceCorrelation} of the observations of $a_{00}$ and $a_{\ell m}$ is investigated. Note that the distance correlation coefficient of two random variables is zero if and only if the random variables  are independent. Thus, the small values of the edcc, shown in Figure~\ref{fig:corr_coef}, justify the assumption of independence. Moreover, the coefficients $a_{00}$ which predominantly influence the size of the particles are described by a log-normally distributed random variable $Z_{00}$ with  parameters $\mu=0.4,\sigma= -29.83$, whose probability density $f_{0}:\R_+\to\R_+$ is given by
\begin{align}
    f_{0}(x\mid \mu,\sigma) &= \frac{1}{\sqrt{2\pi}\sigma x}\exp{\left(  -\frac{(\ln(x)-\mu)^2}{2\sigma^2} \right)}, \qquad\text{for each } x\in \R_+.
\end{align}

\begin{figure}[H]
    \centering
    \includegraphics[width=\textwidth]{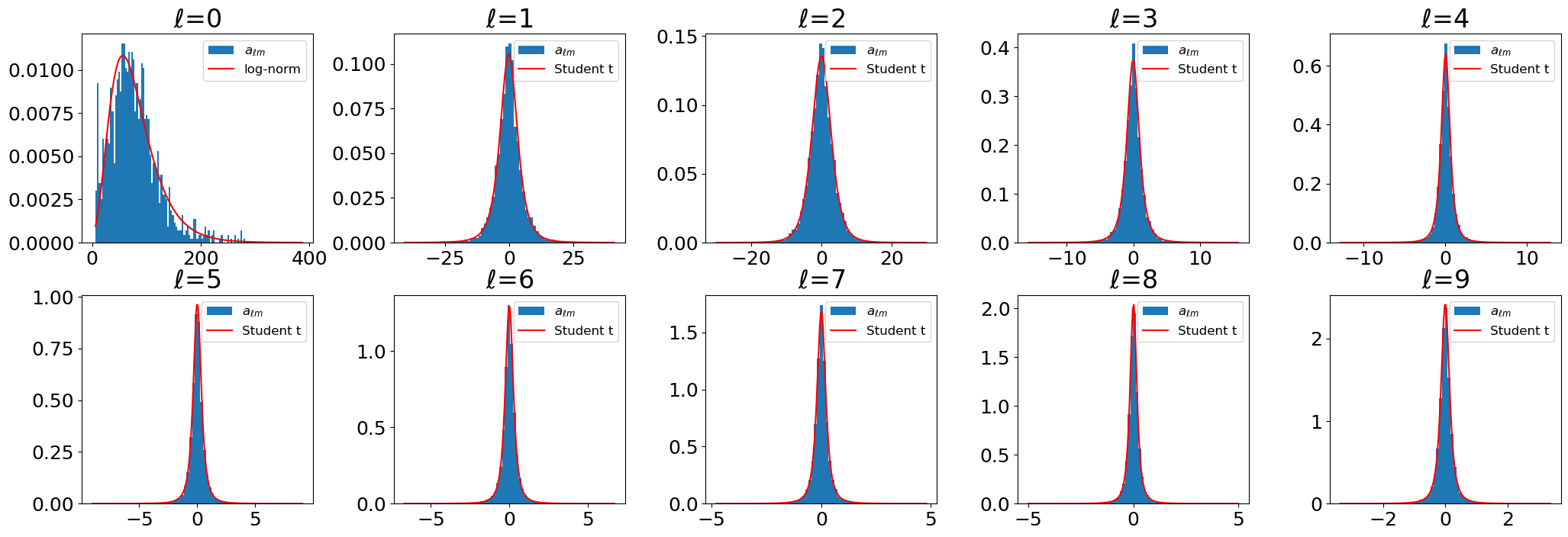}
    \caption{Distribution of spherical harmonics coefficients: The blue histograms depict the distributions of coefficients $a_{\ell m}$ across different orders $\ell$, while the red curves represent the probability densities of the fitted parametric distributions. In particular, for the case where $\ell=0$, a log-normal distribution is fitted, while in the other cases, Student's t distributions are used.}
    \label{fig:sh_coeff}
\end{figure}

On the other hand,  for each $\ell>0$, the coefficients $a_{\ell m}$  are described by a scaled Student's t-distribution, i.e., for some $\nu,\tau>0$, the probability  density $f_{\ell}:\R_+\to\R_+$ of
the random variable 
$\tau Z_{\ell m} $ is given by
\begin{align}
    f_{t}(x\mid\nu)&=\frac{\Gamma(\frac{\nu+1}{2})}{\Gamma(\frac{\nu}{2})\sqrt{\pi \nu }}\left(1+ \frac{x^2}{\nu}\right)^{-(\nu+1)/2}, \qquad \text{for each }x\in \R,
\end{align}
where $\Gamma$ and $\ln$ denote the gamma function and  the natural logarithm, respectively.
 The fitted values of the parameters $\nu$ and $\tau$  are given in Table~\ref{tab:sh_coeff}, see also Figure~\ref{fig:sh_coeff} for a visual validation of the fitted densities $f_{\ell}:\R_+\to\R_+$ for $\ell=0,1,\ldots,9$.

\begin{table}[H]
    \centering
    \begin{tabular}{c|c|c|c|c|c|c|c|c|c}
         parameter $\setminus$ order&$\ell=1$&$\ell=2$&$\ell=3$&$\ell=4$&$\ell=5$&$\ell=6$&$\ell=7$&$\ell=8$&$\ell=9$\\\hline
         $\nu$ & 2.22&4.64&2.84&2.43&2.28&2.26&2.20&2.17&2.22  \\\hline
         $\tau$ & 3.51&2.79&0.98&0.56&0.37&0.28&00.21&0.17&0.15
    \end{tabular}
    \caption{Values of fitted model parameters $\nu$ and $\tau$.  }
    \label{tab:sh_coeff}
\end{table}

The resulting outer shell model $\{X(u),u\in S^2\}$ is given by
\begin{align}
    X(u)=\sum_{\ell=0}^L\sum_{m=-\ell}^\ell Z_{\ell m}Y_{\ell m}(u)\qquad\mbox{for each $u\in S^2$,}
\end{align}
where $Y_{\ell m}$ denotes the  spherical harmonic function introduced in Eq.~\eqref{def.sph.har}. 
Realizations of this model are shown in Figure~\ref{fig:outer_shell_process}f. Note that the model has been calibrated on a voxel grid, see Eq.~(\ref{eq::particle_representation_radius_function}). Thus, the generated outer shells are scaled by the voxel side length of 128~nm afterwards.

\subsection{Stochastic grain architecture model}
\label{sec:GBPD}

This section introduces the  model used to describe the inner grain architecture of NMC811 particles. The model is based on a so-called generalized balanced power diagram (GBPD)~\cite{alpers2015GBPD}, which is a generalization of the well-known Voronoi~\cite{doi:10.1137/Voronoi} and Laguerre~\cite{Lautensack_Zuyev_2008} tessellations. GBPDs are a powerful tool to describe Euclidean space filling structures, such as grains, in a low parametric way~\cite{petrich2021efficient}. This low-parametric representation allows for data compression, fast computation, and a simpler and more interpretable grain architecture modeling as compared to voxel-based approaches.

In general, for any integer $d\in\N=\{1,2,\ldots\}$, a GBPD  is a decomposition of the $d$-dimensional Euclidean space $\R^d$ into non-overlapping  subsets (or grains) $G_1, ..., G_\numSeeds\subset \R^d$, where for each $i\in\{1,\ldots,\numSeeds\}$ it holds that
\begin{align}
G_i=\{x \in \R^d \colon |x-s_i|_{M_i}-r_i\leq|x-s_j|_{M_j}-r_j\; \mbox{for all  $j\in \{1,\ldots,\numSeeds\}\setminus\{i\}$}\}, \label{eq:tessellation}
\end{align}
given $\numSeeds \in \N$ distinct seed points $s_1, ..., s_\numSeeds \subset \R^d$ and markers $(M_1,r_1),...,(M_\numSeeds,r_\numSeeds)$. Here, $M_i\in \R^{d \times d}$ is a positive definite matrix, $r_i\in \R$ is an offset parameter and $|x-s_i|^2_{M_i}=(x-s_i)^\top M_i(x-s_i)$ is the metric induced by $M_i$.   Later on,  for $d \in \{2,3\}$,  each set  $G_i$ will represent a single grain and consequently will be referred to as such. Figures~\ref{fig:2D_fit} and \ref{fig:combine_scales} show examples of such tessellations for the cases $d = 2$ and $d=3$, respectively.


The tessellations used within the proposed grain architecture model are further restricted to GBPDs whose matrices $M_i$ can be written as $M_i=Q_i D_i Q_i ^\intercal$. Here, $Q_i$ is a basis change matrix to a given basis whose first basis vector is $s_i/ |s_i|$ and $D_i$ is a diagonal matrix. This restriction reduces the marker dimensionality from $6+1$ to $3+1$ and allows deriving a non-stationary but radial symmetric model, the realizations of which resemble the grain architectures observed in the image data considered in Section~\ref{sec:descriprion_of_orientation}. To derive such a radial symmetric model, the seed points $s_1, ..., s_\numSeeds$ as well as the markers $(D_1,r_1),...,(D_\numSeeds,r_\numSeeds)$ of the tessellations are modeled stochastically. Thereby, the modeling of the seed points is done by a Poisson point process~\cite{chiu2013stochastic,PoissonLast}, i.e., a random point pattern, whereas the modeling of the markers is accomplished by a fully connected neural network as explained in Section~\ref{sec:tess_fit} below. The neural network based approach allows for a stereological model fitting procedure. This is crucial since only 2D EBSD data are available to investigate the grain architecture of NMC811 particles.


The following section describes this stereological procedure in more detail.

\subsection{Stereological tessellation model calibration}
\label{sec:tess_fit}

The neural network used for marker generation is calibrated by means of a generative adversarial network (GAN) framework~\cite{GANOverview}. Note that GANs represent a game-theory inspired zero-sum game framework, where two key neural networks comprise the GAN framework: The generator $G$ and the discriminator $D$. These neural networks compete with each other to achieve a so-called 
Nash equilibrium~\cite{bacsar1998NE}, i.e., the generator  continually refines its ability to produce data resembling measured samples, while the discriminator  iteratively enhances its capability to distinguish between measured and simulated data. Typically, this is realized by an image generating neural network $G\colon \mathcal{N} \to \mathcal{I}$, i.e., a neural network that maps some noise space $\mathcal{N}$ to some set of  images $\mathcal{I}$, and a discriminator $D\colon \mathcal{I} \to \R$  that is trained to distinguish between measured and simulated images, see ~\cite{SliceGAN,GANOverview} for details. In the present paper, the discriminator tries to assign measured data a label of 1 and simulated ones a label of 0. Using the least squares loss~\cite{Mao_2017_ICCV}, this results in the following min-max problem:
\begin{align}
    \max_{G} \min_{D}\E \left[(D(X)-1)^2 \right] +  \E \left[(D(G(Z)))^2 \right],
    \label{eq:normal_gan_loss}
\end{align}
where $X \in \mathcal{I}$ is a random vector following the distribution of measured data, and $Z \in \mathcal{N}$ is some random vector which serves as input for the generator network $G$ (where $Z$ is sometimes  chosen as white noise~\cite{Mao_2017_ICCV}). The terms $\displaystyle \max_{G}$ and $\displaystyle  \min_{D}$ in Eq.~\eqref{eq:normal_gan_loss} refer to the maximum and minimum taken over all possible parameter values of the neural networks $G$ and $D$, respectively, given that their architectures are fixed. However, in the GAN approach proposed in the present paper, the output of the generator, i.e., a 3D tessellation-based grain architecture, can not  directly be compared to  measured data, i.e., pixel-based 2D image data, and, therefore, the loss function given by Eq.~(\ref{eq:normal_gan_loss}) has to be adapted~\cite{SliceGAN}. Thus, we introduce two (random) functions, $\TReal$ and $\TGen$, which compute (random) discretized 2D cutouts of
2D images drawn from $X$, and 2D cutouts
of simulated 3D tessellations drawn from $G(Z)$, respectively. This leads to the optimization problem
\begin{align}
    \max_G \min_D \E \left[(D(\TReal(X))-1)^2 \right] + \E \left[(D(\TGen(G(Z))))^2 \right].
    \label{eq:adapted_gan_loss}
\end{align}
In the following, the precise definitions of $G, \TReal, \TGen$, and $D$ will be explained in more detail. To ensure an efficient GAN training, we assume the superimposed mapping $D(\TGen(G))$ to be differentiable almost everywhere with respect to the  parameters of $G$. Since $G$ and $D$ are defined by neural networks and thus almost everywhere differentiable, only $\TGen(G)$ has to be analyzed in more detail. To achieve differentiability of $\TGen(G)$, the tessellation representation given in Eq.~(\ref{eq:tessellation}) is not sufficient. Thus, a more general differentiable softmax representation is introduced instead.

\begin{figure}[H]
    \includegraphics[width=\linewidth]{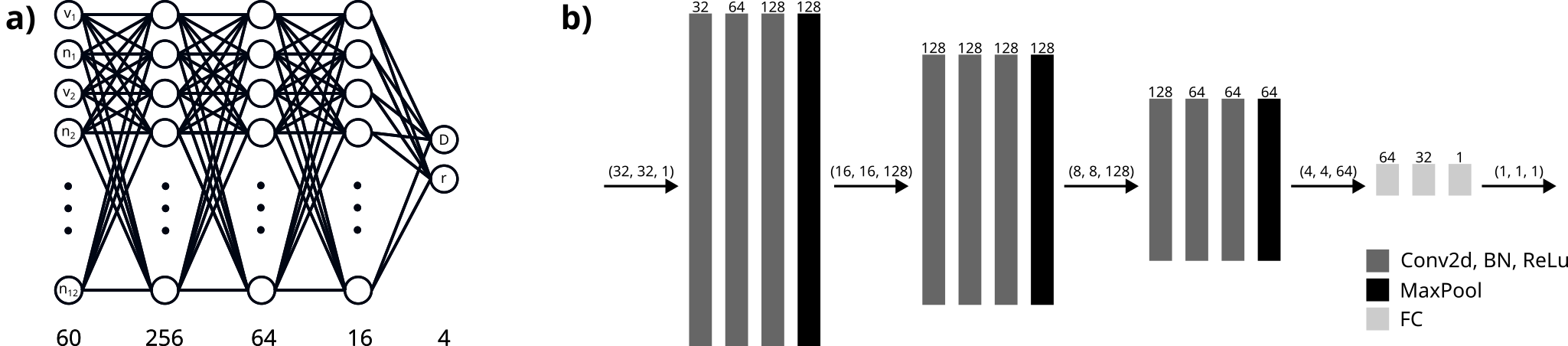}
      \caption{Architectures of neural networks: a) Fully connected architecture of marker generator $\GNet$ being part of $G$. b) Convolutional architecture of discriminator $D$. The numbers above the arrows and bars correspond to the  dimensions of respective inputs/outputs and convolutional layers. 
      }
      \label{fig:architecture}
\end{figure}

\subsubsection{Tessellation generation}\label{sec.tes.gen}
For a given sample $(s_1,n_1,\gamma_1),\ldots,(s_\numSeeds,n_\numSeeds,\gamma_\numSeeds) \in W= [-128,128]^3\times[0,1]^2\times[0,\pi]$ of  seed points $s_1,\ldots, s_\numSeeds$ and  noise $(n_1,\gamma_1),\ldots,(n_\numSeeds,\gamma_\numSeeds)$ , the generator $G$ computes markers $(M_1,r_1),...,(M_\numSeeds,r_\numSeeds)$ and a differentiable softmax representation $\DifTesRep \colon \R^3 \to [0,1]^\numSeeds$ of a tessellation, see~\cite{petrich2021efficient,goodfellow2016deep}. This means that the values  of $\DifTesRep$ are given by 
\begin{align}
    \DifTesRep(x) =  \displaystyle
    \begin{pmatrix}
    \exp{(-|x-s_1|_{M_1}+r_1)}\\
    \vdots\\
    \exp{(-|x-s_\numSeeds|_{M_\numSeeds}+r_\numSeeds)}
    \end{pmatrix}
    \frac{1}{\sum_{i=1}^\numSeeds \exp{(-|x-s_i|_{M_i}+r_i})}\qquad\mbox{for each $x\in\R^3$}.
    \label{eq:differentiabel_repr}
\end{align}
The vector-valued mapping $\DifTesRep=(\DifTesRep_1\ldots,\DifTesRep_{\nu})$ directly implies a tessellation $G_1,\ldots,G_\nu$, as defined in Eq.~(\ref{eq:tessellation}), by considering the index $i$ of the component $\DifTesRep_i(x)$ with maximal value, i.e., $\displaystyle G_i=\{x \in \R^3 \colon \DifTesRep_i(x)=\max\nolimits_j \DifTesRep_j(x)\}$.
Furthermore, note that the differentiable softmax representation $\DifTesRep$ given in Eq.~\eqref{eq:differentiabel_repr} for 3D tessellations can be analogously defined for planar 2D tessellations. 

The generation of the sequence $(s_1,n_1,\gamma_1),\ldots,(s_\numSeeds,n_\numSeeds,\gamma_\numSeeds)$, i.e., the seed points $s_1,\ldots, s_\numSeeds$ and corresponding noise $(n_1,\gamma_1),\ldots,(n_\numSeeds,\gamma_\numSeeds)$ necessary for random marker creation, is done by drawing samples from a homogeneous Poisson point process  with some intensity $\mu>0$, being restricted to the set $W=[-128,128]^3\times[0,1]^2\times[0,\pi]$~\cite{chiu2013stochastic,PoissonLast}. 
However, the corresponding markers  $(M_1,r_1),...,(M_\numSeeds,r_\numSeeds)$  are generated by a fully connected (FC) neural network, which is denoted as $\GNet$. 
Since the grain shape and the size is only influenced by nearby grains, we assume that the marker of a seed point only depends on its neighboring seed points. More precisely, for the generation of the marker $(M_i,r_i)$ only information is considered derived from the  $\consideredNeighbours = 12$ nearest seed points $s_{i_1},\ldots,s_{i_{\consideredNeighbours}}, i_j \in \{1,\ldots,\numSeeds\}\setminus\{i\}$ with respect to the Euclidean distance. Furthermore, to ensure a statistically radial symmetric output of the generator network $\GNet$, its input is modified to get rid of anisotropic information, i.e., the direction of $s_i$. This is done by rotating the coordinate system in such a way that $s_i$ is always aligned in the direction of the unit vector $e_1=(1,0,0)$. That is, for a given seed point $s_i$ we determine a rotation matrix $Q_i$ given by
\begin{align}
    Q_i = R_{e_1,\gamma_i} \cdot R_{\widetilde{s}_i+e_1,\pi}, \label{eq:Q_i}
\end{align}
where $R_{s,\gamma}$ denotes the rotation matrix that describes a rotation around $s\in\R^3$ by an angle of $\gamma\in[0,\pi]$, $\widetilde{s}_i= s_i/ |s_i|$ is the normalized version of $s_i$, and $\gamma_i$ is some angle determined by the Poisson point process. Finally, the vectors $v_{i_1},\ldots,v_{i_\nu}$, which are used as input of $\GNet$ for marker generation, are given by rotating the displacements $s_{i_j}-s_i$ by $Q_i$, i.e., $v_{i_j}=Q_i (s_{i_j}-s_i)$ for $j=1,\ldots, \consideredNeighbours$.
Note that $Q_i$ is a basis change as described in Section~\ref{sec:GBPD}.

To further allow the generator network $\GNet$ to find a non-deterministic grain architecture for given seed points, and therefore being a suitable stochastic model, seed point specific noise $n_{i_j}\in[0,1]^2$ is added. Thus, the input of $\GNet$ is given by $(v_{i_1},n_{i_1},\ldots v_{i_{\consideredNeighbours}},n_{i_{\consideredNeighbours}})$.  Figure~\ref{fig:architecture}a shows the fully connected architecture of $\GNet$. It uses ReLu activation functions $f(t)=\max\{0,t\}, ~ t\in \R$ in the hidden layers for neuron activations and utilizes scaled sigmoid functions $\phi(t)_{a,b}=a+({b-a})/({1+e^{-t}}),~t \in \R$ as  activation functions in the output layers to ensure that the generated markers are in a reasonable range. This limits the elongation of the resulting grains, which makes them less likely to be disconnected (something that can occur for GBPDs), this helps to prevent unintended artifacts, and improves the generator-discriminator training. In particular, the values of $a,b\in\R$ of the channels within the output layer are chosen such that the value of $r_i$ and the diagonal entries of $D_i$ are restricted to $[1,10]$ and $[0,50]$, respectively. A more detailed pseudocode representation of this procedure is shown in Algorithm~\ref{alg:generate_tessellation}.\\

\begin{algorithm}[H]
\caption{Tessellation Generation}\label{alg:generate_tessellation}
\begin{algorithmic}[1]
\Procedure{$G$}{$S = ((s_1,n_1,\gamma_1),\ldots,(s_\numSeeds,n_\numSeeds,\gamma_\numSeeds)) \subset \R^3\times[0,1]^2\times[0,\pi]$}
    \For{$i \text{ in } 1,\ldots,|S|$} 
        \State $Q_{i} \gets R_{e_1,\gamma_i} \cdot R_{\widetilde{s}_i+e_1,\pi}$ \Comment{See Eq.~(\ref{eq:Q_i})}
        \State Compute $i_j,~ j=1,\ldots, \consideredNeighbours$ \Comment{Nearest neighbor indices}
        \State $v_{i_j}\gets Q_i (s_{i_j}-s_i),~ j=1,\ldots, \consideredNeighbours$ \Comment{Relative directions}
        \State $D_i,r_i \gets \GNet(v_{i_1},n_{i_1},\ldots,v_{i_{\consideredNeighbours}},n_{i_{\consideredNeighbours}})$
        \State $M_i \gets Q_i^\top D_i Q_i$ \Comment{Markers}
    \EndFor
    \State \Return $\DifTesRep$ \Comment{See Eq.~(\ref{eq:differentiabel_repr})}
\EndProcedure
\end{algorithmic} 
\end{algorithm}

\begin{figure}[H]
    \centering
    \includegraphics[width=0.8\textwidth]{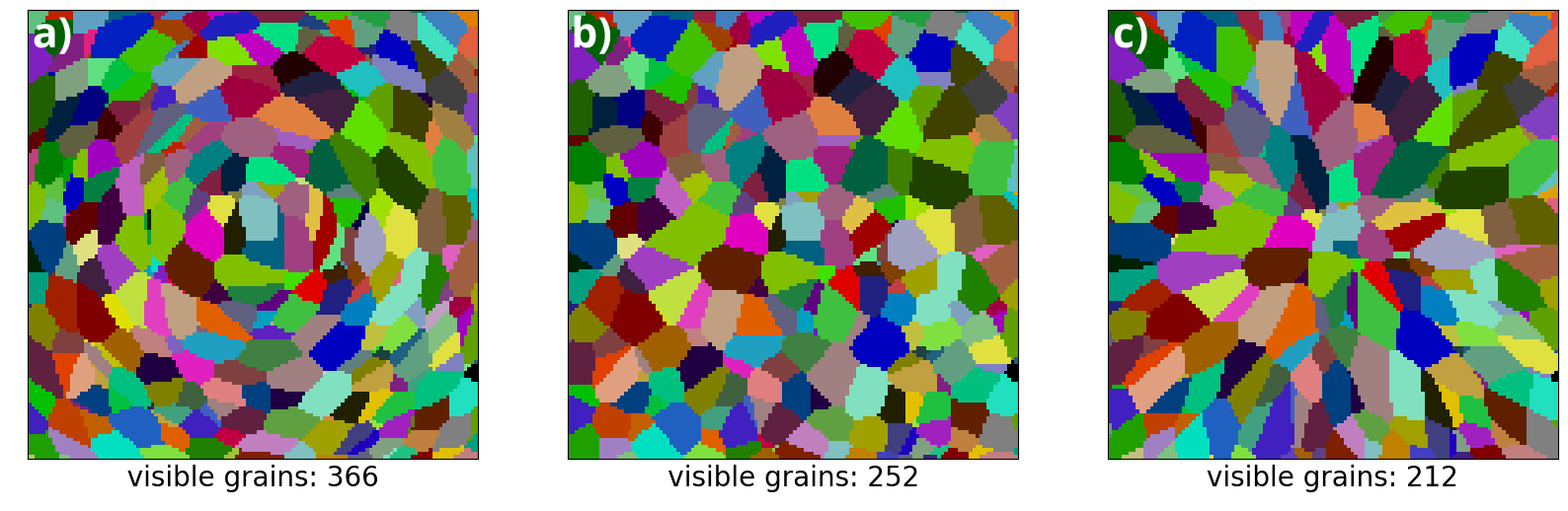}
    \caption{Impact of markers on the grain architecture, where planar sections of 3D grain architectures are depicted with fixed seed points $s_1, \ldots, s_{8000}$, additive marks $r_1, \ldots, r_{8000}$ being equal to 0, and various diagonal marker matrices $D_1, \ldots, D_{8000}$: (a) Diagonal matrices $D_i$ with diagonal entries $(3,1,1)$, (b) Diagonal matrices $D_i$ with diagonal entries $(1,1,1)$, (c) Diagonal matrices $D_i$ with diagonal entries $(1,3,3)$ for each $i \in\{1,\ldots,8000\}$. The number of visible grains reaches from 366 (a), via 252 (b) to 212 (c).}\label{fig::intensity_marker}
\end{figure}

Recall that the sequence 
$(s_1,n_1,\gamma_1),\ldots,(s_\numSeeds,n_\numSeeds,\gamma_\numSeeds) \in W=[-128,128]^3\times[0,1]^2\times[0,\pi]$, 
consisting of seed points $s_i$, noise $n_i$, and uniformly sampled angles $\gamma_i$, is generated by a homogeneous Poisson point process with some intensity $\mu>0$. 
This point process corresponds to the noise generation random variable $Z$ considered in Eq.~(\ref{eq:adapted_gan_loss}). 
A Poisson-distributed random variable is used to determine the number $\numSeeds$ of sampling points.  Then, the locations of the respective points are independently and uniformly sampled within the window $W =[-128,128]^3\times[0,1]^2\times[0,\pi]$. Note that this kind of point process has only one parameter, the intensity $\mu$, i.e., the expected number of points per unit volume. Since each of these points correspond to a grain, the parameter $\mu$ has to be fitted such that the average numbers of visible grains  of generated and measured data in planar 2D cutouts match. However, this cannot be accomplished a priori, since the number of visible grains in a 2D planar section of a 3D tessellation is influenced by the number of seed points as well as their markers, as can be seen in Figure~\ref{fig::intensity_marker}.  To overcome this problem, the intensity $\mu$ is adopted iteratively while training, i.e., the intensity $\mu_{e+1}$ of the Poisson point process at training epoch $e+1,~ e\in \N$ is defined as follows:
\begin{align}
    \mu_{e+1}=
    \begin{cases}
    \mu_e\left(0.5+0.5\frac{\numberGrainsGT}{\numberGrainsGenerated}\right), & \text{ if }e>1,\\
    0.0018, &\text{ if }e=1,  
    \label{eq::intensity}
    \end{cases}
\end{align}
where $\numberGrainsGT,\numberGrainsGenerated >0$
correspond to the average number of observable grains per unit area in planar 2D cutouts of ground truth data and generated data at training epoch $e$, respectively. To avoid large jumps of $\mu_e$, which would hinder the learning process of both generator and discriminator, the intensity $\mu_{e+1}$ is chosen as a convex combination of the previous intensity $\mu_e$ and the desired intensity $\mu_e~(\numberGrainsGT/\numberGrainsGenerated)$. The initial value of this  procedure is set heuristically to  $\mu_1=0.0018$.

As mentioned above, the output of the generator $G$, i.e., $\DifTesRep$ as defined in Eq.~(\ref{eq:differentiabel_repr}), can not be directly compared to measured EBSD data. One reason for this is the circumstance that there are several pixel positions in the EBSD data to which no grain is assigned at all.
Moreover, measured data is only present as  planar 2D sections, therefore planar sections have to be extracted from generated 3D grain architectures to ensure comparability.

\subsubsection{Data transformation} \label{sec:data_transformation}

To address the issue of unassigned pixels, a differentiable representation $\DifTesRep$ (see Eq.~(\ref{eq:differentiabel_repr})) of the restricted GBPD with $d=2$ is fitted to the EBSD data using the method described in~\cite{petrich2021efficient}. This fitted 2D tessellation is referred to as ``ground truth data''. Figure~\ref{fig:2D_fit} shows an exemplary fit. The approach of considering a 2D tessellation fit  to the experimental EBSD  data as ground truth data has at least three advantages. First, it provides a meaningful prescription for missing measurement information (inner white pixels in Figure~\ref{fig:2D_fit}, left). Second, it ensures that the planar sections with which the generator attempts to imitate can indeed be generated, i.e., the ground truth data can be represented by GBPDs.  Without assurance that the ground truth data can be represented by GBPDs, it is possible that the discriminator differentiates between measured and simulated data solely based on a difference of the representation, which would offer no insightful feedback to guide the generator's improvement. Third, it facilitates effective data augmentation during training, i.e., by modifying the seed points and markers of the fitted 2D tessellation slightly, similar but new 2D grain architectures can be generated as additional training data. Note that the distribution of the random variable $X$ considered in Eq.~(\ref{eq:adapted_gan_loss}) describes the distribution of 2D training data, i.e., the data that arise from the ground truth data by this kind of augmentation.

\subsubsection{2D representation of cutouts}

Despite the fact that now both ground truth and simulated data are represented as differentiable tessellations, their dimensionalities are not yet consistent (being equal to $2$ and $3$, respectively). To overcome this problem and to generate input that is feasible for the discriminator $D$, i.e., a matrix of fixed size, randomly located (2D) pixel-based cutouts of these tessellations are computed.

\begin{figure}[H]
    \includegraphics[width=0.85\textwidth]{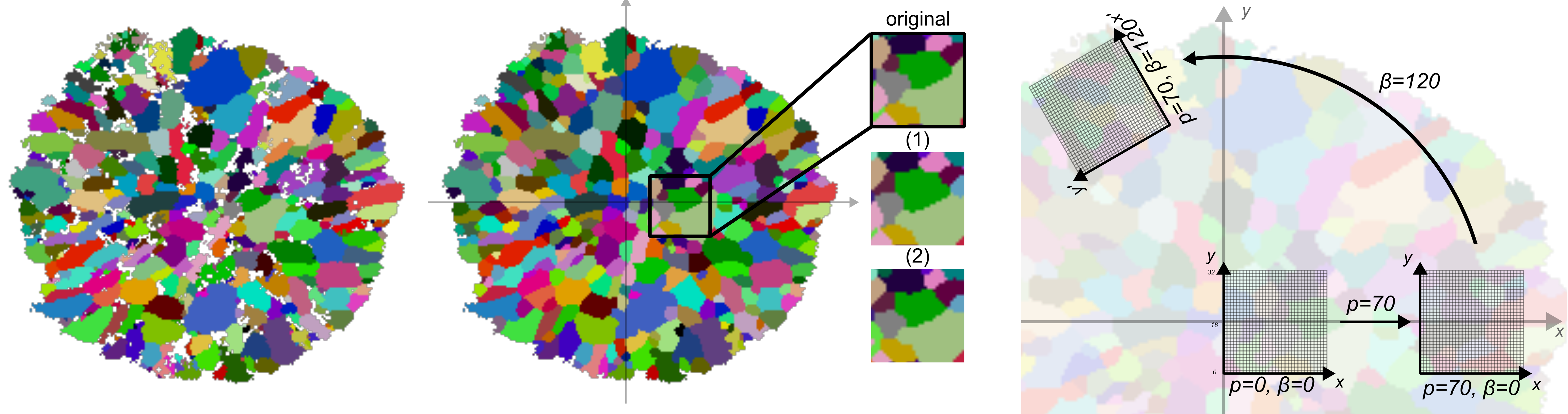}
    \caption{ Raw EBSD image (left) and  correspondingly fitted 2D tessellation with 
    values assigned for missing pixels (middle). Additionally, three exemplary cutouts are magnified, one from the fitted tessellation and two further cutouts, denoted by (1) and (2), arising from data augmentation, see Sections~\ref{sec:data_transformation} and~\ref{sec:training}. Furthermore, sketch of the procedure of  pixel-based cutout computation as described above  (right).
     }
    \label{fig:2D_fit}
\end{figure}

For this purpose, a $32 \times 32$ grid of equidistant points is chosen, say  $\{1,\ldots,32\}\times\{1,\ldots,32\}\subset\R^2$. The tessellation $\DifTesRep$ of the restricted GBPD with $d=2$ is then evaluated at these grid points, resulting in a matrix $I$ with  $32 \times 32 \times \numSeeds$ entries. Such a matrix is often referred to as a multichannel 2D image, with dimensions representing the  $x$-axis, $y$-axis, and channels, respectively. For each $(x,y,i)\in\{1,\ldots,32\}^2\times\{1,\ldots,\nu\}$, the entry  $I_{x,y,i}$ of $I$  corresponds to the probability that the grid point $(x,y)$ belongs to grain $i$. To reduce the number of channels of  $I$ to a (sufficiently small) fixed value, the set of channels is truncated to those channels which belong to grains being most present among the considered grid points.
In the following,  for the case of a 2D tessellation $\DifTesRep$, this procedure is described in more detail.
First, the originally chosen $32 \times 32$ grid  $\{1,\ldots,32\}^2$ of equidistant points in $\R^2$ 
is shifted and rotated at random, i.e.,
a (random)  grid of equidistant points is determined by drawing a radial distance $p$ and a polar angle $\beta$ uniformly from the intervals $(0,80]$ and $[0,2\pi]$, respectively. Then, for  each $(x,y)\in\{1,\ldots,32\}^2$, the corresponding grid point $(x^\prime,y^\prime)$ of the transformed grid is given by
\begin{align}
  (x^\prime,y^\prime)=  R_\beta \binom{x+p}{y-16.5},\quad \mbox{where $R_\beta = \begin{pmatrix}
    \cos{\beta} & -\sin{\beta}\\
    \sin{\beta} & \cos{\beta}
\end{pmatrix}$}.\label{gri.poi.pri}
\end{align}
Evaluating the tessellation $\DifTesRep$ at these points gives the (non-truncated) image $I=(I_{x,y,i})$, where the entry $I_{x,y,i}$ of the matrix $I$ at $(x,y,i)\in\{1,\ldots,32\}^2\times\{1,\ldots,\nu\}$ is given by
\begin{align}
     I_{x,y,i} = \DifTesRep_i \left(R_\beta \binom{x+p}{y-16.5}\right).\label{eq:cutout_2D}
\end{align}
The right-hand side of Figure~\ref{fig:2D_fit} shows the three steps of this cutout computation procedure are illustrated, where two different  coordinate systems are used, one for the original pixel positions $(x,y)\in\{1,\ldots,32\}^2$ and one for the grid points $(x',y')$ introduced in Eq.~\eqref{gri.poi.pri}. The procedure starts with a cutout, whose center lies on the $x$-axis, i.e., $p=\beta=0$. Grains located in this cutout that are strongly orientated towards the center of the planar section (i.e., $\alpha<< \frac{\pi}{2}$, see Eq.~\eqref{eq:orientation}) are elongated in the direction of the $x$-axis. In a second step, the  cutout is shifted in the direction of $x$-axis, according to the selected value of $p$ (where $p=70$ in Figure~\ref{fig:2D_fit}). As before, grains within the cutout which have an orientation $\alpha<< \frac{\pi}{2}$ show elongations in the direction of the $x$-axis of the cutout. Finally, in  a third step, the  cutout is rotated by an angle of $\beta\in[0,2\pi]$. This does not only displace the cutout but also changes its orientation. Consequently, grains that are located in this cutout having an orientation of $\alpha<< \frac{\pi}{2}$, are  orientated in the direction of the (new) $x^\prime$-axis of the cutout. This procedure is performed to generate consistent cutouts which are feasible for the discriminator $D$ and, simultaneously, to be able to use a wide range of cutout positions within the ground truth data for training.

To achieve a fixed (small) number of channels necessary for the discriminator input, the $\numSeeds$ channels of $I$ are sorted by the sum of their entries in descending order. Then, all but the first 12 channels are omitted, i.e, only the channels corresponding to the 12 most prominent grains regarding the grid points are retained, in order to retain the channels that contain the most information of grain morphologies. This (random) procedure will be denoted by $\TReal$ in the following.

A similar procedure, analogous to that described above  for $\TReal$, is used to determine $32\times32\times 12$  matrices which represent (random) 2D cutouts of the simulated 3D tessellations introduced in Section~\ref{sec.tes.gen}, for a given number $\nu$ of seed points. More precisely, for a uniformly drawn radial distance $p \in ( 0,80]$, a matrix $I=(I_{x,y,i})$ with
  $32 \times 32 \times \numSeeds$ entries is computed, where 
\begin{align}
     I_{x,yi} = \DifTesRep_i \left(\begin{pmatrix}
         x+p\\y-16.5\\0
     \end{pmatrix}\right) \qquad\mbox{for each $(x,y,i)\in\{1,\ldots,32\}^2\times\{1,\ldots,\nu\}$}.\label{eq:cutout_3D}
\end{align}
Due to the rotational invariance of both the Poisson point process of seed points and the marker generation, and the ability to create infinity many simulated grain architectures, no rotation matrix is needed in Eq.~\eqref{eq:cutout_3D} for the generation of grain locations, unlike the  procedure for $\TReal$ described above.
The resulting matrix $I$ (a multichannel 2D image) is then channelwise sorted by the sum of their entries and cropped to maintain a constant number of channels, using a procedure similar to the one already described. This (random) procedure of  computing $32\times32\times 12$  multichannel 2D images from artificially generated 3D tessellations is denoted by $\TGen$.
Recall that both, $\TReal$ and $\TGen$ pay attention to differentiability and to preserve a consistent orientation of the cutout with respect to the origin of the tessellation. This is crucial to enable the discriminator to capture preferred grain orientations with respect to the particle center.

\subsubsection{Discriminator architecture} 

The discriminator $D$, which is trained to decide whether a $32\times32\times12$ cutout image originates from ground truth or artificially generated data, is structured as a convolutional neural network (CNN). Note that CNNs are a  neural networks that can handle spatial dependencies in their input and are therefore especially useful in computer vision tasks such as image classification~\cite{alzubaidi2021review}. To avoid overfitting issues, the discriminator $D$ considered in the present paper has a rather simple architecture. Namely, the discriminator consists of stacked convolutional layers, batch normalization layers, and ReLU activation functions. These components are responsible for feature extraction, learning acceleration, as well as neuron activation and deactivation. Additionally, max-pooling layers are used for dimension reduction, and dense layers are used to process the extracted features to a single output. Figure~\ref{fig:architecture}b shows a detailed representation of the network, and the number of features per layer.

\subsubsection{Training procedure and data augmentation} \label{sec:training}

\begin{figure}[H]
    \centering
    \includegraphics[width=.8\textwidth]{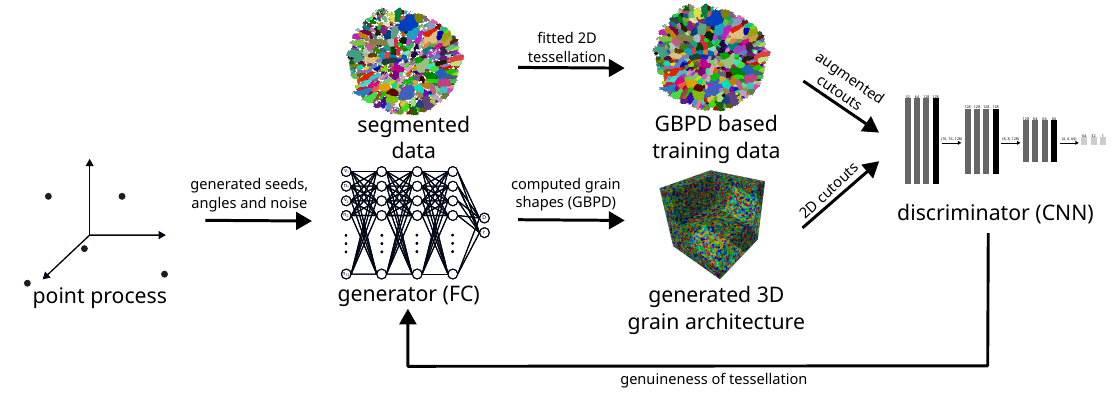}
    \caption{GAN training procedure to generate artificial grain architectures.}
    \label{fig:procedure}
\end{figure}

The GAN-based training process involves iteratively training the generator $G$ and discriminator $D$ through gradient-descent and a non-gradient-descent based optimization of the point process intensity $\mu$, as described in Eq.~(\ref{eq::intensity}). The gradient-descent based training uses an Adam optimizer with a learning rate of $10^{-4}$ and gradient normalization~\cite{mandic2004generalized}. The whole training procedure is done over 200 epochs with 100 steps per epoch and a batch size of 64 and 128 for generator and discriminator training, respectively. While training the discriminator the pixel values of its input images (see Eqs.~(\ref{eq:cutout_2D}) and~(\ref{eq:cutout_3D})) are rounded to the closest value in $\{0,1\}$ to suppress features that arise from the dimensions of the underlying artificially generated and ground truth tessellations, respectively.  To address the issue of overfitting the discriminator, the update of the discriminator's weights is skipped during training, provided that its current mean squared error given by Eq.~(\ref{eq:adapted_gan_loss}) is below $0.3^2$. Furthermore, training data augmentation is achieved through marker modification, i.e., the diagonal entries of the matrices $D_i$ introduced in Section~\ref{sec:GBPD},
and the additive marks $r_i$ of the fitted 2D tessellation  are modified through uniform augmentations up to $20\%$, i.e., $\frac{|x-x'|}{|x|}<0.2$, where $x$ is equal to $r_i$ or a diagonal entry of $D_i$, and $x'$ is its augmented version. This  augmentation of training data results in more diverse grain architectures that still adhere to the constraints of the restricted GBPD representation, see Figure~\ref{fig:2D_fit}. However, this is contrary to augmentations achievable through conventional techniques of image data processing~\cite{10.1145/3510413}. Figure~\ref{fig:procedure} and 
Algorithm~\ref{alg:training_procedure}  provide an overview of the training procedure described above, where the pseudocode is presented in such a way as to improve readability and should not be considered computationally efficient. For a more efficient implementation, the seed point generation  and, thus, the 3D grain architecture generation, as well as the 2D grain architecture augmentation can be restricted to a small sampling window that contains the observed planar section.

\begin{algorithm}[H]
\caption{Training Procedure}\label{alg:training_procedure}
\begin{algorithmic}[1]
\Procedure{Train}{}
\State $lr \gets 0.0001$ \Comment{Learning rate}
\State $bs \gets 128$  \Comment{Batch size}
\State $\mu_1 \gets 0.0018$ \Comment{Point process intensity}
\For{$e \text{ in } 1,\ldots,200$} \Comment{Epochs}
    \For {$step$ in $ 1,\ldots,100$}
        \State Draw $S_1,\ldots,S_{bs/2}$ from $Z$ with intensity $\mu_{e}$ \Comment{Seeds, noises and angles }
        \State $\mathrm{loss}\gets \frac{2}{bs} \sum_{i}^{bs/2} (D(\TGen(G(S_i))))^2$ 
        \State Maximize $\mathrm{loss}$ with respect to $\GNet$ and $lr$ \Comment{One step of Adam gradient descent}
             \State Draw $S_1,\ldots,S_{bs}$ from $Z$ with intensity $\mu_{e}$ \Comment{Seeds, noises and angles }
            \State Draw $x_1,\ldots,x_{bs}$ from $X$ \Comment{Augmented measured planar sections}
            \State $\mathrm{loss}\gets \frac{1}{bs} \sum_{i}^{bs} (D(\round(\TReal(x_i)))-1)^2 +  D(\round(\TGen(G(S_i))))^2$ \Comment{$\round(a) = \underset{ b\in \N^{32\times 32 \times 12}}{\argmin} |a-b|$ (rounding)}
        \If {$\mathrm{loss} > 0.09$} \Comment{Avoid discriminator overfitting}
            \State Minimize $loss$ with respect to $D$ and $lr$ \Comment{One step of Adam gradient descent}
        \EndIf
        
        \State  $\mu_{e+1}\gets\mu_e(0.5+0.5\frac{\numberGrainsGT}{\numberGrainsGenerated}),$ \Comment{Iterative intensity adaption, see Eq.(\ref{eq::intensity}})
    \EndFor
\EndFor
\EndProcedure
\end{algorithmic} 
\end{algorithm}

\subsection{Multi-scale model} 

\begin{figure}[H]
\begin{minipage}{0.45\textwidth}
    \centering
    \includegraphics[width=\textwidth]{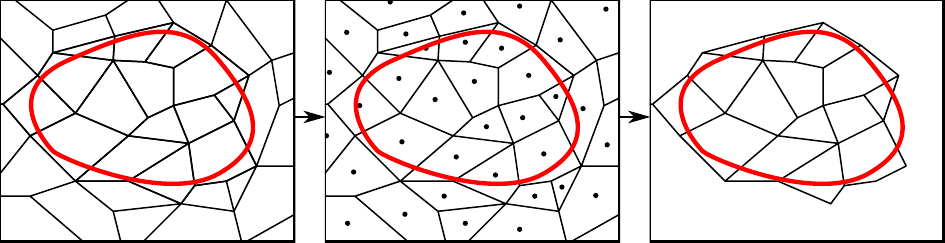}
\end{minipage}
\hspace{12pt}
\begin{minipage}{0.45\textwidth}
    \centering
    \includegraphics[width=.9\textwidth]{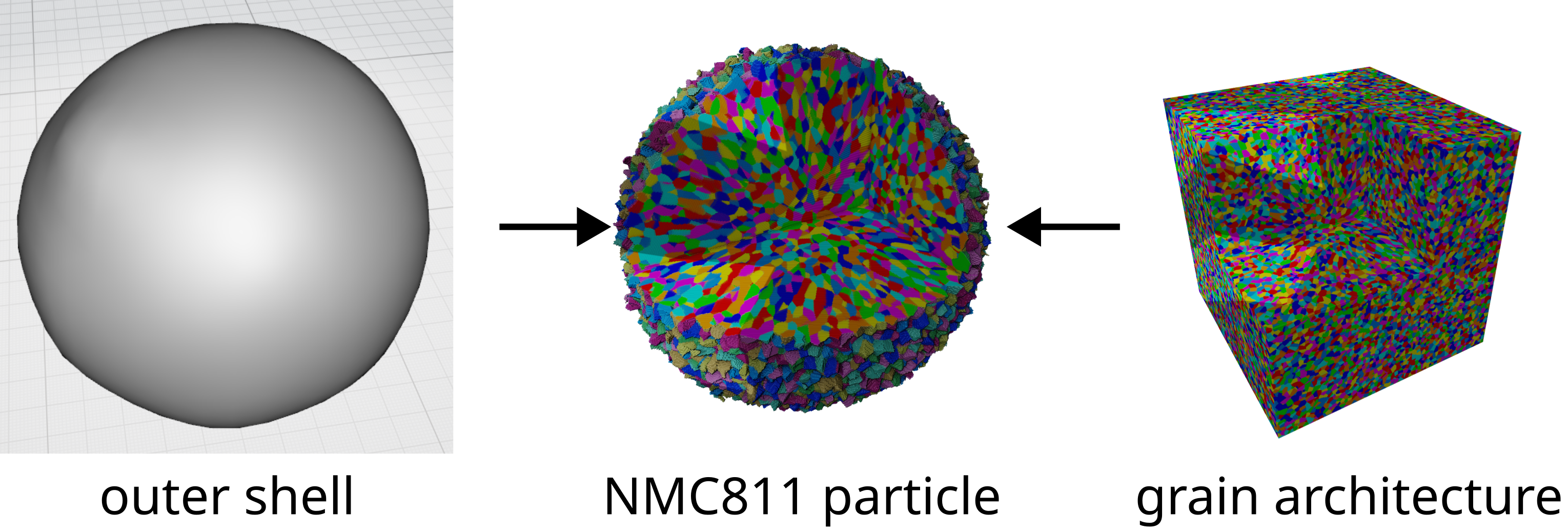}
\end{minipage}
    \caption{2D sketch of the overlay and grain removing procedure (left). The black dots inside the grains correspond to their centers of mass. Grains with centers of mass outside the originally generated outer shell (red) are removed. 
    Additionally, 3D renderings of samples are presented, derived from both the outer shell model and the grain architecture model, as well as their combination into a virtual NMC811 particle (right).
    }
    \label{fig:combine_scales}
\end{figure}

Combining the outer shell model and the grain architecture model follows the  approach presented in~\cite{FuratMultiScale}. More specifically, to get a simulated NMC811 particle, an outer shell and a grain architecture are  independently drawn from the models introduced in Sections~\ref{sec:outer_shell} 
and \ref{sec:GBPD}, respectively, and
 overlaid on a 3D domain.  Next, grains whose centers of mass are not located inside the outer shell are removed. 
Figure~\ref{fig:combine_scales} illustrates this procedure and shows  3D renderings of samples drawn 
    from the outer shell model and the grain architecture model,  as well as their combination to a virtual NMC811 particle. 

\section{Results and discussion}
\label{sec:res_and_disc}

In this section, the simulated NMC811 particles are evaluated quantitatively, where various structural descriptors, characterizing particle and grain morphology, are computed for ground truth and simulated data and, afterwards, compared to each other.  Importantly, the structural  descriptors considered in this section have not been used for model calibration.

\subsection{Validation of the outer-shell model}

To evaluate the outer-shell model introduced in Section~\ref{sec:outer_shell}, four different descriptors are chosen to characterize the 3D morphology of ground truth and simulated microstructures, respectively.  These descriptors include the  diameter $\diameter$,  the compactness $\compactness$,  the aspect ratio $\aspectRatio$, and the  sphericity $\sphericity$ of particles~\cite{compactnesssphericity}. More precisely,  for a particle represented by its radius function  $P\colon S^2 \to  \R_+$, the values of these descriptors are given by
\begin{align}
    \diameter(P) =& \max \{P(u)+P(-u) \colon u \in S^2 \},\\
    \compactness(P) =& \frac{V(P)}{V(\partial \text{Conv}(P))},\\
    \aspectRatio(P) =& \frac{\diameter(P)}{\minDiameter(P)}, \quad\mbox{where $  \minDiameter(P) = \min \{P(u)+P(-u) \colon u \in S^2 \}$},\\
    \sphericity(P) =& \frac{\pi^{\frac{1}{3}}(6V(P))^{\frac{2}{3}}}{A(P)}, \label{eq::sphericity}
\end{align}
where $A(\cdot)$ is the surface area, $V(\cdot)$ the enclosed volume, and $\partial \text{Conv}(\cdot)$ the boundary of the convex hull.  Note that the compactness $\compactness$ measures the degree of convexity of a particle, whereas the aspect ratio $\aspectRatio$ describes the particle's elongation, and the sphericity $\sphericity$ measures the roundness of a particle, i.e., its similarity to a sphere. Figure~\ref{fig:outer_shell_descriptors}~(top row) shows histograms of these descriptors for both ground truth and simulated particles.

In the outer-shell analysis, simulated particles  having a volume of less than $10$ voxels, i.e., less than 10 $\cdot$ (0.128~$\upmu$m)$^3$, are neglected. These  volumes are omitted for further analysis, since the segmentation procedure described in Section~\ref{sec:segmentation_and_labeling} generates only particles with larger sizes. The histograms of the descriptors   $\diameter$,  $\compactness$, $\aspectRatio$, and  $\sphericity$ of the resulting particles show nice agreement  with correspnding histograms computed from nano-CT data, see Figure~\ref{fig:outer_shell_descriptors}.
It is noteworthy that the model does not produce particles with extremely low sphericities, as observed in the segmentation process (see the values for $\sphericity \leq 0.6$ in Figure~\ref{fig:outer_shell_descriptors}). Further investigations are required to check whether these particles can be attributed to an imperfect segmentation, and consequently, whether their absence in model realizations is important. 

Recall that the relatively small edcc values displayed in Figure~\ref{fig:corr_coef} provided the basis for assuming that the model's random spherical harmonics coefficients are independent of each other. However, an increasing trend of the edcc along with  the order of the basis functions can be observed. It is assumed that this trend can partially be attributed to missing fine structures in particles with a small diameter due to the discrete voxel-based resolution. Additionally, the small expected values of $|Z_{\ell m}|$ for $ \ell>4$ suggest that these coefficients can be neglected, without decreasing the model quality significantly. A systematic investigation of the influence of the maximum spherical harmonics order $L$ on the model quality will be the subject of future work.

\subsection{Validation of the grain architecture model}

\begin{figure}[H]
    \centering
    \includegraphics[width = \textwidth]{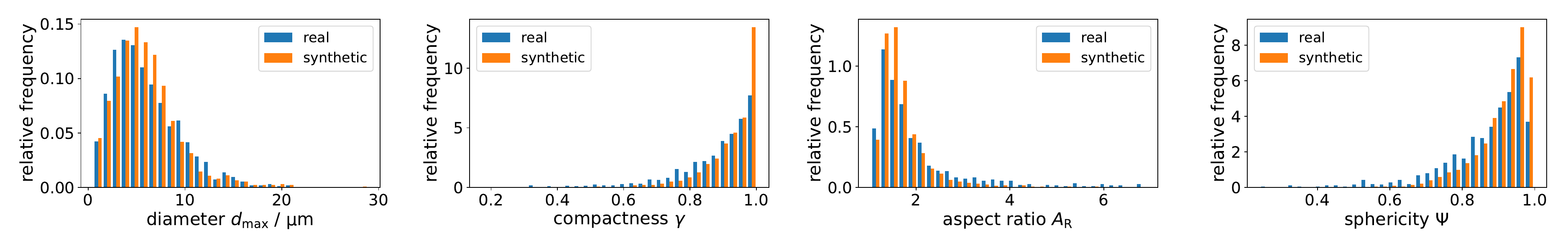}

    \includegraphics[width = \textwidth]{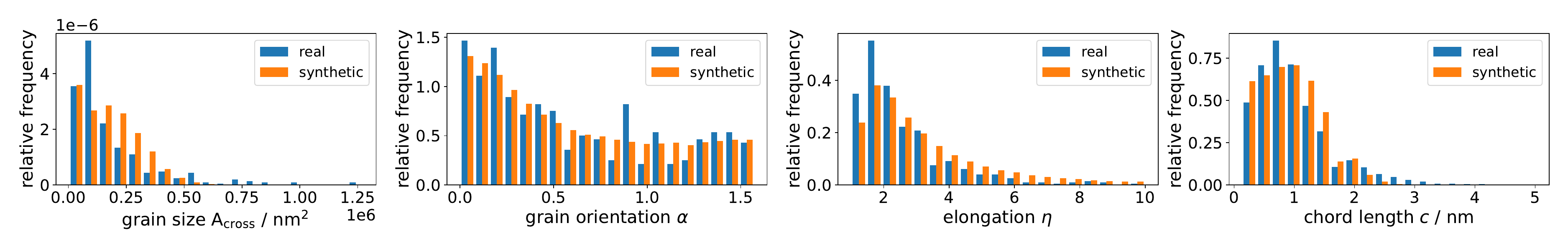}
    \caption{ Histograms for ground truth data (blue) of outer shells (top row) and inner grain architecture (bottom row) of NMC811 particles,  compared with histograms computed for   simulated data drawn from the respective models (orange). The 3D descriptors $\diameter$,  $\compactness$, $\aspectRatio$, and  $\sphericity$ characterizing the outer shell are computed by means of the radius function representation given in Eq.~(\ref{eq::particle_representation_radius_function}), whereas the four 2D descriptors of the inner grain architecture (bottom row) are computed by means of  pixel-based planar sections.}
     
    \label{fig:grain_descriptors}
     \label{fig:outer_shell_descriptors}
\end{figure}

The model introduced in Section~\ref{sec:GBPD}
for the inner 3D grain architecture of NMC811 particles is based solely on 2D EBSD data. Since only 2D planar section data of ground truth grain architectures is available, the validation of simulated grain architectures is done by a statistical comparison of pixel-based planar sections. For this comparison, the distributions of four microstructure descriptors are evaluated:  the distribution of the size $\grainSize$ of planar grain sections, the distributions of their  elongation $\elongation$ and  their orientation $\orientation$, and  the chord-length distribution~\cite{chordlength} of the ensemble of planar grain sections. Note that the size $\grainSize$ of a planar grain section is given by the number of pixels occupied by the given grain section. The  orientation $\orientation$ is defined in Eq.(\ref{eq:orientation}), and the  elongation $\elongation$ is given by
\begin{align}
    \elongation &= \frac{e_1}{e_2},
\end{align}
where $e_1 \geq e_2>0$ are the eigenvalues of the principal components of the planar section of the given grain, see Section~\ref{sec:descriprion_of_orientation}. The chord-length distribution is given by the distribution of the lengths $\chordLength$ of all chords of the planar section, where  a chord of a pixel-based planar section of grains is a sequence of consecutive aligned pixels that belong to the same grain and that can not be extended further without containing pixels belonging to other grains.
Figure~\ref{fig:grain_descriptors}~(bottom row) shows histograms of these descriptors for both ground truth and simulated data, which are are quite similar to each other. However,   the elongation $\elongation$ shows slightly larger values for simulated data, which means that the grains observed in the  planar  sections of the stereologically fitted model are slightly more elongated than those in the ground truth data. A similar behavior can be observed for the grain orientation $\orientation$, where simulated grain architectures show a slightly more preferred orientation as compared to the ground truth. Nevertheless, 
the histograms computed from ground truth and simulated data, respectively,
show nice agreement, which indicates that   planar 2D sections of grains observed in EBSD data are well captured by the stochastic grain architecture model introduced in Section~\ref{sec:GBPD}.  

\subsection{Discussion of model assumptions}

\begin{figure}[H]
    \centering
    \includegraphics[width =.3 \textwidth]{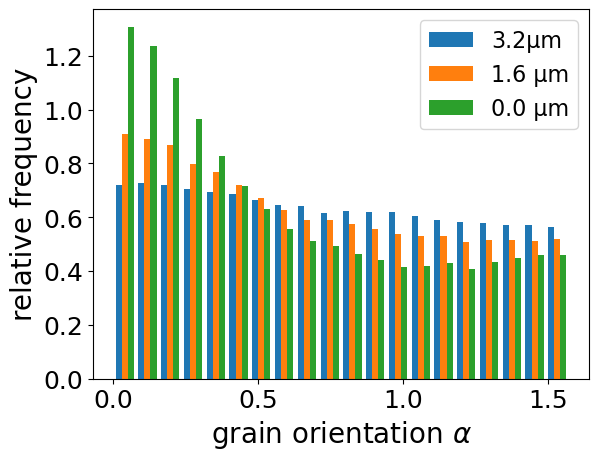}
    \caption{Histograms of grain orientations for different cut heights.
    Planar sections with a larger distance to the particle center show less preferred grain orientation.}
    \label{fig:orientation_cut_height}
\end{figure}

The trained grain architecture model, which was fitted and evaluated by means of  planar 2D sections, can  be used to generate realistic  3D grain architectures of NMC811 particles and to investigate structural properties of these 3D morphologies. For example, Figure~\ref{fig:orientation_cut_height}  shows how the distribution of grain orientations observed in planar sections depends on the section's position (i.e., its cut height). It can be observed that the grains that are visible in planar sections near the origin exhibit an increasingly preferred orientation, in contrast to more distant planar sections, which show an almost uniform distribution of grain orientations. This observation coincides with the assumption stated in Section~\ref{sec:descriprion_of_orientation} that some planar sections observed in EBSD data may not pass through the particle center and therefore were neglected for training purposes. However, the GAN-based training of the grain architecture model could be adapted such that planar sections that are taken farther away from the particle center can also be used. Nevertheless, such modifications would necessitate providing the discriminator with additional information about the observed data. Such additional information could include the distance of the planar section to the particle center. This adjustment would enable the discriminator to evaluate the plausibility of observed grain elongations or preferred orientations based on the position of the planar section within the particle.  However, this type of information, i.e., the relative location of the planar section with respect to the particle center, would have to be acquired by laboratory measurements.

In the proposed multi-scale model, we assumed that the inner grain architecture can be modeled independently of the outer shell. While this may be true for large particles, it will not be true for particles with very small radii, i.e., particles that consist only of hundreds of grains. To investigate this issue further and, consequently, adapt the modeling approach, EBSD data of differently sized particles has to be acquired. Importantly, these rather small particles account for a small mass percentage of an electrode and, thus, their inclusion/exclusion is not expected to significantly influence predicted electrode performance.

\section{Conclusion}
The multi-scale modeling approach described here enables realistic 3D particle geometries with sub-particle grain detail. The 3D sub-particle grain morphologies were generated by only using 2D EBSD data, facilitating characterization of 3D grain morphological properties from 2D experimental information. Representative particle generation capabilities are not only required for characterizing otherwise challenging 3D morphological properties but are also required for high-fidelity cathode degradation models that require significant test geometries to mitigate stochastic effects resulting from particle geometry variations. Notably, it is significantly easier and faster to simulate chemo-mechanics on hundreds of generated geometries, as compared to relying solely on a sparse set of empirical image data. A bulk of such simulated NMC811 particles with inner grain architectures, suitable for numerical simulations, is made publicly available. Previous cathode degradation modeling efforts focused mostly on NMC532 particles with randomly distributed grains, which was primarily due to the availability of full 3D imagery~\cite{ALLEN2021230415}. Now, with the capability of inferring 3D geometries of grain architectures with curved grain boundaries from only 2D EBSD slices, it is possible to model NMC811 degradation and consider radially oriented grain geometries, which will be the focus of future work. Additionally, the proposed method can enable quicker development of generation methods for new and emerging electrode chemistries.

In the future, there are plans to expand the model not only to capture the geometry of the grains but also to describe their crystallographic orientation. Additionally, there are intentions to adjust the fitting for the outer shell model such that calibration can be performed from 2D image data to achieve a fully stereological multi-scale model.

\section*{Acknowledgements}

This work has been partially supported by the German Research Foundation (DFG) under grants 673729 and 678514. This work was authored in-part by Alliance for Sustainable Energy, LLC, the manager and operator of the National Renewable Energy Laboratory for the U.S. Department of Energy (DOE) under Contract No. DE-AC36-08GO28308. Funding was provided by DOE's Vehicle Technologies Office, Extreme Fast Charge and Cell Evaluation of Lithium-ion Batteries Program, Jake Herb, Technology Manager. The views expressed in the article do not necessarily represent the views of the DOE or the U.S. Government.

\section{Data availability}
 The datasets generated during and/or analyzed during the current study are available from the corresponding authors on reasonable request.
 
 \section{Code availability}
 All formulations and algorithms necessary to reproduce the results of this study are described in the “Stochastic multi-scale model” section.

\bibliographystyle{unsrt}
\bibliography{refs,refs_pw}

\end{document}